\documentclass[%
 reprint,
 amsmath,amssymb,
 aps,
]{revtex4-2}

\usepackage{physics,graphicx,xcolor}
\usepackage{colortbl}
\usepackage[most]{tcolorbox}

\usepackage{dsfont}

\begin{document}


\title{Random density matrices: Analytical results for mean root fidelity and the mean-square Bures distance}

\author{Aritra Laha}
\author{Agrim Aggarwal}
\author{Santosh Kumar}
 \email{skumar.physics@gmail.com}
\affiliation{%
 Department of Physics, Shiv Nadar University
Gautam Buddha Nadar, Uttar Pradesh 201314, India
}%


\begin{abstract}
Bures distance holds a special place among various distance measures due to its several distinguished features and finds applications in diverse problems in quantum information theory. It is related to fidelity and, among other things, it serves as a bona fide measure for quantifying the separability of quantum states. In this work, we calculate exact analytical results for the mean root fidelity and mean square Bures distance between a fixed density matrix and a random density matrix, and also between two random density matrices. In the course of derivation, we also obtain spectral density for the product of the above pairs of density matrices. We corroborate our analytical results using Monte Carlo simulations. Moreover, we compare these results with the mean square Bures distance between reduced density matrices generated using coupled kicked tops and find very good agreement.
\end{abstract}

\pacs{Valid PACS appear here}
\maketitle


\section{Introduction}

Various distance measures between quantum states play a crucial role in quantum information theory and find applications in diverse problems, such as quantum communication protocols, quantification of quantum correlations, quantum algorithms in machine learning, and quantum-state tomography~\cite{NC2000,BZ2017,W2017,ZE2011,TBCL2019,STM2013,RSI2016,Gao2016,SO2013,VPRK1997,ACSZC2019,LTOCC2019,KLPCSC2019,CPCC2019,LZ2004,STM2013,KKF2020,WPSW2020,PSW2020}. Roughly speaking, distance measures quantify the degree of \emph{closeness} of two given quantum states. Some examples of prominent distance measures are trace distance, Hilbert-Schmidt distance, Bures distance, and Hellinger distance. Among these, Bures distance holds a special place due to its several notable features. For instance, it is Riemannian and also a monotone. The latter feature signifies that it does not grow under an arbitrary completely positive trace-preserving quantum operation. Due to this monotonicity, among other things, the Bures distance serves as a bona fide measure of distinguishability between two quantum states~\cite{MMPZ2008,PPZ2016,RSI2016}. The Bures distance was introduced by Bures during his investigations pertaining to the infinite products of von Neumann algebras~\cite{B1969}. It was further studied by Uhlmann in his works related to the transition probability of states~\cite{U1976,U1986}. 

Given two quantum states represented by the respective density matrices $\rho_1$ and $\rho_2$, the Bures distance is defined by~\cite{B1969,U1976,BZ2017,W2017},
\begin{equation}
d_{B}(\rho_1,\rho_2)=\sqrt{2-2\sqrt{\mathcal{F}(\rho_1,\rho_2)}},
\end{equation}
where
\begin{equation}
\mathcal{F}(\rho_1,\rho_2)=\left(\tr \sqrt{\sqrt{\rho_1}\rho_2 \sqrt{\rho_1}}\right)^{2}
\end{equation}
is the fidelity~\cite{B1969,U1976,U1986,J1994},
with ``$\tr$" representing the trace. Although not obvious from the above definition, Bures distance, like any other distance measure, is symmetric in its argument, i.e., $d_{B}(\rho_1,\rho_2)=d_{B}(\rho_2,\rho_1)$. It assumes values from the interval $[0,\sqrt{2}]$. The minimum value of 0 is obtained when the two states coincide and the maximum value of $\sqrt{2}$ is achieved when the two states are supported on orthogonal subspaces.

In the course of investigation of properties exhibited by certain quantum states under consideration, it is quite often desirable to have some reference states for comparison. In this regard, random states constitute a natural choice and provide the most typical or generic characteristic~\cite{W1990,H1998,ZS2001,SZ2004,ZPNC2011,CN2016}. They can be associated with quantum states affected by noise. Furthermore, they relate to quantum chaotic systems where one examines states whose classical analogues are chaotic~\cite{HKS1987,W1990,H2010}. Considering finite-dimensional Hilbert spaces, as far as pure states are concerned, there is a natural measure associated with them, viz. the Haar measure on the unitary group. However, for mixed states, there is no such unique measure~\cite{W1990,H1998,ZS2001,SZ2004}. One extremely popular and useful probability measure is the so-called Hilbert-Schmidt measure. This probability measure on the set of finite-dimensional mixed states is realized via the operation of partial trace on a bipartite system~\cite{H1998,ZS2001}, as is described in section~\ref{SecMSB} below. Hilbert-Schmidt probability measure is also induced by the Hilbert-Schmidt distance~\cite{H1998,ZS2001}. Due to its fundamental importance, investigation of these Hilbert-Schmidt distributed random density matrices has remained a very active area of research and many important results related to their various aspects have been worked out~\cite{ZS2001,SZ2004,Lubkin1978,LP1988,Page1993,G2007,MBL2008,NMV2011,KP2011,VPO2016,KSA2017,K2019,FK2019,W2020,CN2016,ZPNC2011}. An example is the celebrated Page formula which gives the average von Neumann entropy associated with a bipartite system~\cite{Page1993}. For distance measures also, there have been some noteworthy progress in the context of random density matrices~\cite{B1996,ZS2005,HLW2006,M2007,PPZ2016,BSZW2016,MZB2017,K2020,KC2020}, however much remains to be explored.

In a recent work~\cite{K2020}, one of the present authors considered the Hilbert-Schmidt distance $d_\mathrm{HS}(\rho_1,\rho_2)=\sqrt{\tr(\rho_1-\rho_2)^2}$ and derived exact analytical expressions for its mean square between a fixed density matrix and a random density matrix, and also between two random density matrices. These results serve as useful references for comparing Hilbert-Schmidt distance between quantum states and are of relevance in applications such as construction of entanglement witness operators~\cite{WPSW2020,PSW2020}, variational hybrid quantum-classical algorithms in machine learning~\cite{TBCL2019,LTOCC2019,KLPCSC2019,CPCC2019}, precision quantum-state tomography~\cite{ZE2011,STM2013,KKF2020}, etc. Unfortunately, as far as distinguishability criterion is concerned, Hilbert-Schmidt distance has its limitations since it is not a monotone~\cite{NC2000,BZ2017,W2017,O2000,WS2009}. In the present work, using random matrix theory (RMT) techniques we calculate exact analytical results for the mean-square Bures distance, 
\begin{align}
\label{DB2}
\langle d^{2}_{B}(\rho_1,\rho_2)\rangle \equiv D_{B}^2(\rho_1,\rho_2)=2-2\langle \sqrt{\mathcal{F}(\rho_1,\rho_2)}\,\rangle,
\end{align}
where the averaging $\langle\cdot\rangle$ is with respect to the probability measure followed by the random density matrices involved. Similar to Ref.~\cite{K2020}, we consider the above quantity for a fixed density matrix and a random density matrix, and also for two random density matrices. For the former case, the Laplace transform approach helps us to reformulate the problem in terms of semicorrelated Wishart random matrix. On the other hand, for the latter, it allows us to establish the connection with the statistics of the product of two Wishart-Laguerre random matrices. In the course of derivation, we also calculate the spectral density of the product of a fixed density matrix and a random density matrix, and also of two random density matrices. 

The rest of the paper is organized as follows. In Sec.~\ref{SecMSB} we derive exact results for the mean root fidelity and mean square Bures distance between the pairs of density matrices as described above. These results are corroborated using Monte Carlo simulations. In Sec.~\ref{SecKT} we compare our analytical results for mean square Bures distance with those computed from coupled kicked top systems. Section~\ref{SecSum} is devoted to summary and outlook. Details of some of the calculations appear in the Appendices.

\section{Mean root fidelity and Mean square Bures distance for random density matrices} 
\label{SecMSB}

Consider a random bipartite pure state $\ket{\psi}$ taken from an $nm$-dimensional Hilbert space $\mathcal{H}_n\otimes\mathcal{H}_m$, where $\mathcal{H}_n$ and $\mathcal{H}_m$ are the $n$- and $m$-dimensional Hilbert spaces associated with the constituent systems. Without loss of generality, we take $n\le m$ and consider the random reduced state $\rho$ induced by the operation of partial tracing over the $m$-dimensional constituent, viz.,
\begin{equation}
\label{rho}
\rho=\frac{\tr_{m}(|\psi\rangle\langle\psi|)}{\langle\psi|\psi\rangle}.
\end{equation}
This $n$-dimensional reduced density matrix $\rho$ is distributed according to the Hilbert-Schmidt probability measure with the corresponding probability density function (PDF) given by~\cite{ZS2001},
\begin{equation}
\label{Prho}
\mathcal{P}(\rho)=C (\det \rho)^{m-n}~\delta(\tr \rho-1)\Theta(\rho).
\end{equation}
Here, the notation $\det(\cdot)$ stands for determinant and $\tr(\cdot)$ is the trace as already indicated earlier. The Dirac delta function $\delta(\cdot)$ in the above equation constrains the density matrix $\rho$ to have unit trace, and the Heaviside theta function $\Theta(\cdot)$ enforces the positive semi-definiteness condition. The normalization factor is given by
\begin{equation}
C=\Gamma(nm)\Big(\pi^{n(n-1)/2}~\prod_{j=1}^{n}\Gamma(m-j+1)\Big)^{-1}.
\end{equation}
As per the random matrix theory, the above reduced density matrix belongs to the fixed-trace Wishart-Laguerre ensemble and has the construction~\cite{ZS2001,SZ2004},
\begin{equation}
\label{map}
\rho=\frac{W}{\tr W},
\end{equation}
where $W$ is a matrix taken from the (unconstrained) Wishart-Laguerre ensemble with PDF,
\begin{equation}
\label{PW}
P(W)=(\Gamma(nm))^{-1}C\, (\det W)^{m-n}e^{-\tr W}\Theta(W).
\end{equation}
It should be noted that the PDF of $\rho$, as in Eq.~\eqref{Prho}, can be obtained using that of $W$ as
\begin{equation}
\mathcal{P}(\rho)=\int d[W] P(W)\,\delta\left(\rho-\frac{W}{\tr W}\right),
\end{equation}
where the flat measure $d[W]$ is defined in terms of the matrix elements of $W$ as $d[W]=\prod_{i}dW_{ii}\cdot\prod_{j<k}d\, \mathrm{Re}(W_{jk})\,d\, \mathrm{Im}(W_{jk})$. As can be seen, the above equation essentially normalizes the Wishart-Laguerre random matrix $W$ to have unit trace and thereby gives the distribution of the random density matrix $\rho$.

In the following, we evaluate the mean root fidelity and hence the mean-square Bures distance between: (A) a fixed density matrix $\sigma$ and a random density matrix $\rho$, and (B) two independent random density matrices $\rho_1$ and $\rho_2$ which, in general, may differ in the dimension of the traced-out subsystem; say $m_1$ and $m_2$. In the process, we also compute the spectral density of the product matrices $\tau=\sqrt{\sigma} \rho \sqrt{\sigma}$ and $\chi=\sqrt{\rho_1} \rho_2 \sqrt{\rho_1}$. We note that since density matrices are Hermitian as well as positive semi-definite, the matrices $\tau$ and $\chi$ are also so because of their symmetrized product structure. Moreover, the matrices $\rho\sigma$ (or $\sigma\rho$) and $\rho_1\rho_2$ (or $\rho_2\rho_1$), although not Hermitian in general, share the same eigenvalues as $\tau$ and $\chi$, respectively.

\subsection{A fixed density matrix $\sigma$ and a random density matrix $\rho$}
\begin{figure*}[!t]
\includegraphics[width=1.0\linewidth]{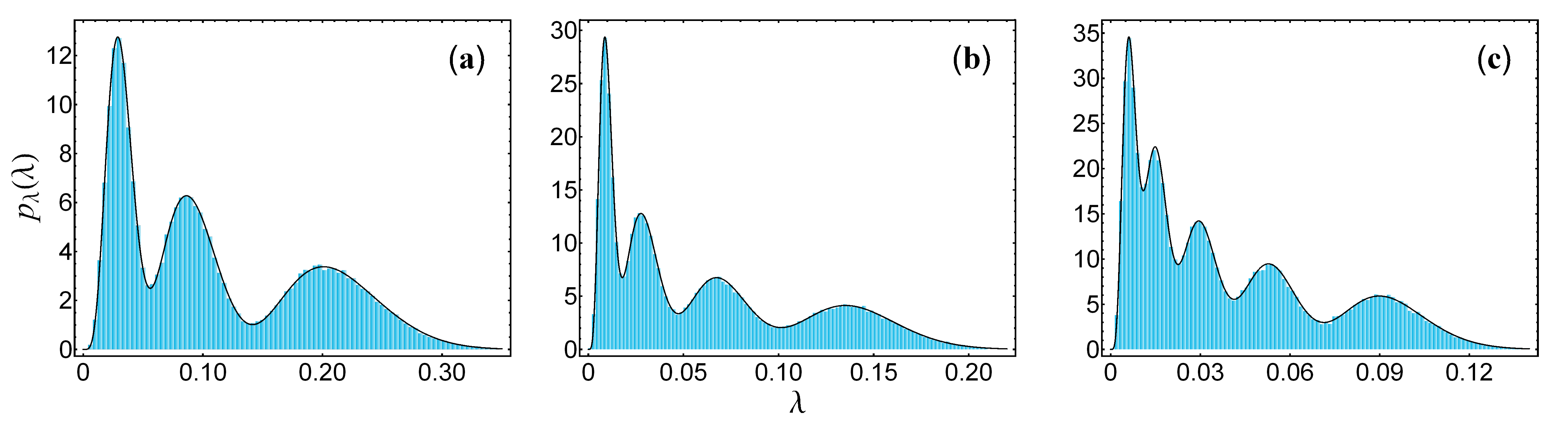}
\centering
\caption{Plots of the spectral density of the matrix $\tau=\sqrt{\sigma}\rho \sqrt{\sigma}$. The dimensions $(n,m)$ and eigenvalues $(1/a_1,...,1/a_n)$ of $\sigma$ are (a) $(3, 8)$ and $(0.15,0.33,0.52)$; (b) $(4, 9)$ and $(0.07,0.17,0.35,0.41)$; and (c) $(5, 10)$ and $(0.09,0.12,0.21,0.28,0.30)$. The solid curves represent analytical results and the histograms are based on simulations.}
\label{fig1}
\end{figure*}

\subsubsection{Spectral density of $\tau=\sqrt{\sigma} \rho \sqrt{\sigma}$}

In this case only one of the density matrices is random. Therefore, we can evaluate the PDF for the product matrix $\tau$ as,
\begin{equation}
\label{ptaua}
\mathcal{P}_\tau(\tau)= \int \delta(\tau -\sqrt{\sigma}\rho\sqrt{\sigma})~\mathcal{P}(\rho)~d[\rho],
\end{equation}
where $d[\rho]$ is defined similar to $d[W]$. As demonstrated in the Appendix~\ref{AppA}, the above can be written in terms of an inverse Laplace transform ($s\to t=1$) over an integral involving the Wishart-Laguerre density~\eqref{PW},
\begin{align}
\label{ptaub}
\nonumber
&\mathcal{P}_\tau(\tau)=\mathcal{L}^{-1}\Big[\Gamma(nm) s^{-nm}\\
&\times \int\delta(\tau -s^{-1}\sqrt{\sigma}W \sqrt{\sigma})P(W)~d[W]\Big]_{t=1},
\end{align}
with $s>0$. The above equation provides a strategy to calculate the statistics of $\tau$ using the statistics of $\widetilde{W}=\sqrt{\sigma}\,W \sqrt{\sigma}$. The latter is identified as a semicorrelated Wishart random matrix with covariance matrix equal to $\sigma$ and having its PDF proportional to $(\det \widetilde{W})^{m-n}e^{-\tr(\sigma^{-1}\widetilde{W})}$~\cite{G1963}. 

The joint PDF $P_{x}(\{x\})$ for the unordered eigenvalues $x_1,...,x_n$ of $\widetilde{W}$ is known to be~\cite{ATLV2004,SMM2006}
\begin{align}
\nonumber
&P_{x}(\{x\})= \frac{(-1)^{n(n-1)/2}}{n!}\left(\prod_{r>l}\frac{x_r-x_l}{a_r-a_l}\right)\\
&\times\left(\prod_{i=1}^n \frac{a_i^m x_i^{m-n}\Theta(x_i)}{\Gamma(m-i+1)}\right)\det[e^{-a_jx_k}]_{j,k=1}^{n},
\end{align}
 where $a_j^{-1}$'s are the eigenvalues of the matrix $\sigma$ (i.e. $a_j$'s are the eigenvalues of $\sigma^{-1}$). The fact that only the eigenvalues of $\sigma$ appear in this expression is a consequence of the invariance of the distribution of $W$ under unitary conjugation, which implies that $\sqrt{\sigma}\,W \sqrt{\sigma}$ is statistically equivalent to $A^{-1/2}W A^{-1/2}$, where $A=\mathrm{diag}(a_1,...,a_n)$. Now, since the joint PDF $P_{x}(\{x\})$ is symmetric with respect to the eigenvalues, we can examine the behavior of a generic eigenvalue by integrating out $n-1$ eigenvalues and obtaining its marginal density~\cite{Mehta2004,Forrester2010}. If $x$ and $\lambda$ be the generic eigenvalues of $\widetilde{W}$ and $\tau$, with their respective marginal probability densities $p_x(x)$ and $p_\lambda(\lambda)$, then it follows from Eq.~\eqref{ptaub} that
\begin{align}
\label{pl1}
\nonumber
p_{\lambda}(\lambda)&=\mathcal{L}^{-1}\left[\Gamma(nm) s^{-nm}\int_{-\infty}^\infty\!\! \delta(\lambda-s^{-1}x)~p_{x}(x)dx\right]_{t=1}\\
&=\mathcal{L}^{-1}\left[\Gamma(nm) s^{1-nm}~p_{x}(s\lambda)\right]_{t=1}.
\end{align}
Now, the marginal density $p_x(x)$ is known to be~\cite{ATLV2004,SMM2006,RKG2010},
\begin{align}
\label{px}
p_{x}(x)=\frac{1}{n}\prod_{r>l}\frac{1}{(a_r-a_l)}~\sum_{i=1}^{n}\det[\zeta^{(i)}_{jk}]_{j,k=1}^n,
\end{align}
where
\begin{align}
\label{zetaijk}
\zeta^{(i)}_{jk}=\begin{cases}\frac{a_j^m x^{m-i} e^{-a_j x}\Theta(x)}{\Gamma(m-i+1)}, &  k=i,\\
 a_{j}^{k-1}, &  k\ne i.
\end{cases}
\end{align}
We use Eq.~\eqref{px} in Eq.~\eqref{pl1} and push the factor $s^{1-nm}$ inside the $i$th column of the matrix $[\zeta^{(i)}_{jk}]$ and then the Laplace variable $s$ occurs only in this $i$th column. We also note that $\Theta(s\lambda)=\Theta(\lambda)$ for $s>0$. Consequently, the Laplace inversion can be conveniently carried out using the standard result $\mathcal{L}^{-1}[s^{-\alpha}e^{-\beta s}]_t=(t-\beta)^{\alpha-1}\Theta(t-\beta)/\Gamma(\alpha)$. Therefore, we obtain the desired expression as
\begin{equation}
\label{pl2}
p_{\lambda}(\lambda)=\frac{\Gamma(nm)}{n}\prod_{r>l}\frac{1}{(a_r-a_l)}~\sum_{i=1}^{n}\det[\eta^{(i)}_{jk}]_{j,k=1}^n,
\end{equation}
where
\begin{align}
\eta^{(i)}_{jk}=\begin{cases}\frac{a_j^m \lambda^{m-i} (1-a_j\lambda)^{i+nm-m-2}\Theta(\lambda)\Theta(1-a_j\lambda)}{\Gamma(m-i+1)\Gamma(i+nm-m-1)}, & k=i,\\
 a_{j}^{k-1}, &  k\ne i.
\end{cases}
\end{align}
If there are multiplicities in the eigenvalues of $\sigma$, we obtain $0/0$ form. In such cases, analytical results can be obtained from Eq.~\eqref{pl2} using a limiting procedure, for example using L'H\^ospital's rule. 

Of special interests are the cases when the fixed density matrix represents either a pure state or a maximally mixed state. In the former case, one of the eigenvalues of $\sigma$ is 1 and the rest are 0's. In the latter case, all eigenvalues are $n^{-1}$ and correspondingly $\sigma=n^{-1}\mathds{1}_n$. The spectral densities for these cases can be obtained from Eq.~\eqref{pl2} using limiting procedures, however, it is much more straightforward to derive them directly, as done below.\\

\noindent
\emph{Pure state $\sigma$}:
Based on our discussion regarding the invariance of the distribution of $W$ under unitary conjugation, it is evident that the only nonzero eigenvalue of the matrix $\sqrt{\sigma}\,W \sqrt{\sigma}$ equals one of the diagonal elements of $W$. Moreover, in this case, the nonzero eigenvalue of the matrix $\tau=\sqrt{\sigma}\rho\sqrt{\sigma}$ coincides with the fidelity itself. Therefore, we can write
\begin{equation}
\mathcal{F}=\left(\tr\sqrt{\sqrt{\sigma}\frac{W}{\tr W}\sqrt{\sigma}}\right)^2=\frac{W_{11}}{\sum_{k=1}^n W_{kk}}.
\end{equation}
The diagonal elements $W_{kk}$ of $W$ are independent and identically distributed as Gamma random variables with the PDF (see, e.g., \cite{KC2020}), 
\begin{equation}
\label{gamma}
p_w(w)=\frac{1}{\Gamma(m)}w^{m-1}e^{-w}\Theta(w)\,\equiv \text{Gamma}(m).
\end{equation}
Further noticing that $R=\sum_{k=2}^n W_{kk}$ is distributed as Gamma$(nm-m)$, the fidelity $\mathcal{F}=W_{11}/(W_{11}+R)$ turns out to be distributed as Beta$(m,nm-m)$, i.e.,
\begin{align}
\label{pFpure}
p_\mathcal{F}(\mathcal{F})=\frac{\Gamma(n m)\mathcal{F}^{m-1}(1-\mathcal{F})^{nm-m-1}}{\Gamma(m)\Gamma(nm-m)}\Theta(\mathcal{F})\Theta(1-\mathcal{F}).
\end{align}
Alternatively, we can apply the Laplace-inversion trick to obtain the PDF of the fidelity, viz. $p_\mathcal{F}(\mathcal{F})=\mathcal{L}^{-1}\left[\Gamma(nm) s^{1-nm}~p_w(s\mathcal{F})\right]_{t=1}$. This again leads to Eq.~\eqref{pFpure}, as it should. The above expression for fidelity distribution is in complete agreement with the expression obtained by \.Zyczkowski and Sommers using a different albeit equivalent approach~\cite{ZS2005}.\\
\begin{figure}[!t]
\includegraphics[width=0.8\linewidth]{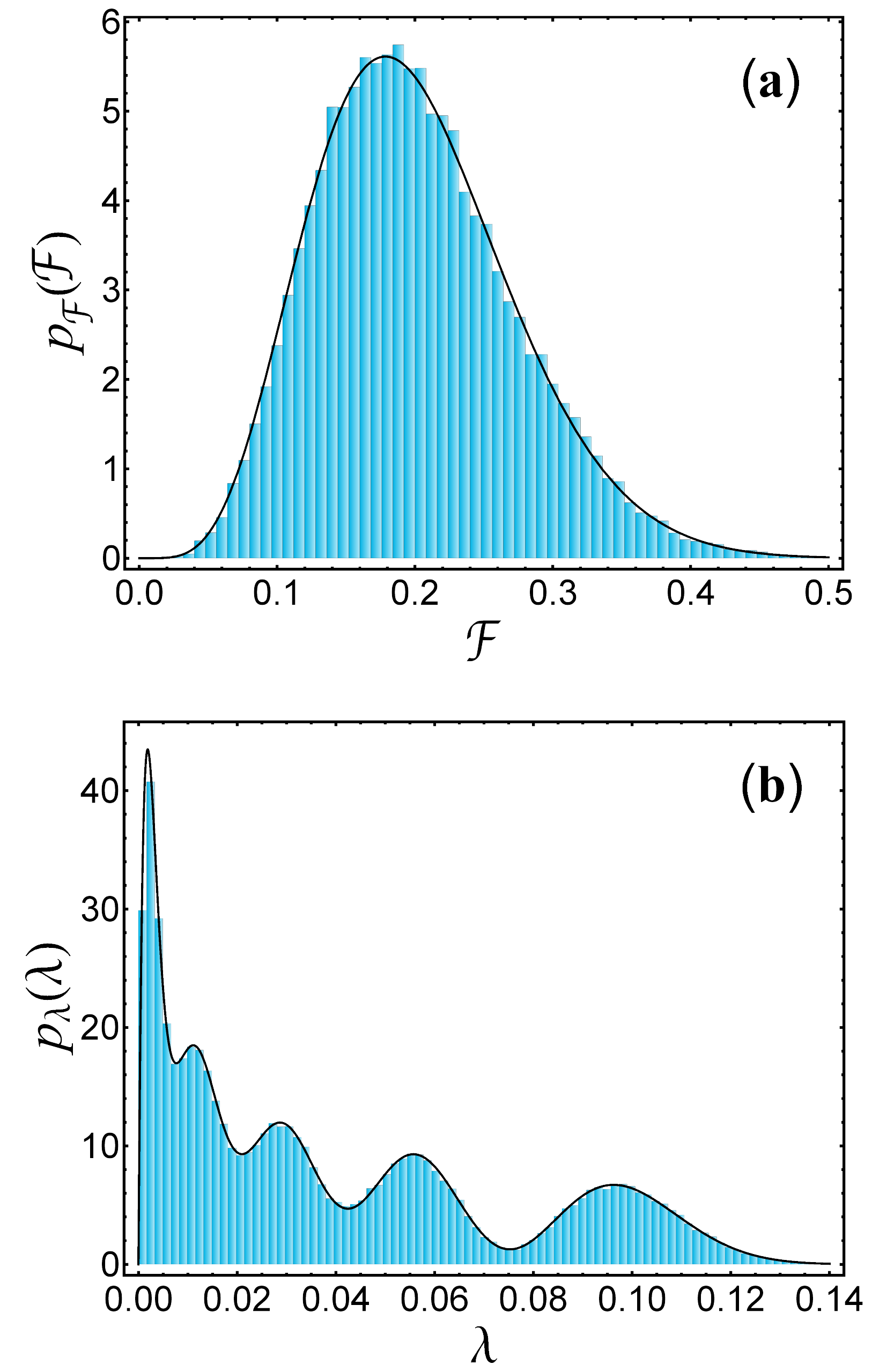}
\centering
\caption{Plots of probability density of (a) fidelity between $\rho$ and $\sigma$ with latter a pure state, (b) a generic eigenvalue of  $\tau=\sqrt{\sigma}\rho \sqrt{\sigma}$ with $\sigma$ corresponding to maximally mixed state. The random density matrix $\rho$ has parameters $n=5$ and $m=6$ in both cases. The solid lines are based on analytical results and the histograms are using simulations.}
\label{fig2}
\end{figure}
\noindent
\emph{Maximally mixed state $\sigma$}:
In this case, as $\sigma=n^{-1}\mathds{1}$, the eigenvalues of $\sqrt{\sigma}W \sqrt{\sigma}$ will be $n^{-1}$ times the eigenvalues of $W$. As such, the marginal density of a generic eigenvalue of $\sqrt{\sigma}W \sqrt{\sigma}$ follows using standard result for Wishart-Laguerre eigenvalue density. Similarly, the eigenvalues of $\tau=\sqrt{\sigma}\rho\sqrt{\sigma}$ will be $n^{-1}$ times the eigenvalues of $\rho$. The distribution of the marginal density of a generic eigenvalue of $\rho$ is already known (see, e.g., \cite{KP2011}), which leads us to
\begin{align}
\nonumber
&p_\lambda(\lambda)=\sum_{i=1}^{n} c_i \,(n\lambda)^{i+m-n-1}(1-n\lambda)^{-i+n m-m+n-1}\\
\nonumber
&\times\big[(n-i)F_{m-n+1}^{-n,i-nm+m-n}-nF_{m-n+1}^{1-n,i-nm+m-n}\big]\\
&\times
\Theta(\lambda)\Theta(1-n\lambda).
\end{align}
Here, $F^{\alpha,\beta}_\gamma:=\,_2F_1(\alpha,\beta;\gamma;\frac{n\lambda}{n\lambda-1})/\Gamma(\gamma)$ with $_2F_1(\cdots)$ being the Gauss hypergeometric function. The coefficient $c_i$ is given by
\begin{align}
c_i=\frac{(-1)^i\,\Gamma(m+1)\Gamma(nm)}{\Gamma(i)\Gamma(n-i+1)\Gamma(i+m-n+1)\Gamma(nm-m+n-i)}.
\end{align}
\begin{figure*}[!t]
\includegraphics[width=1.0\linewidth]{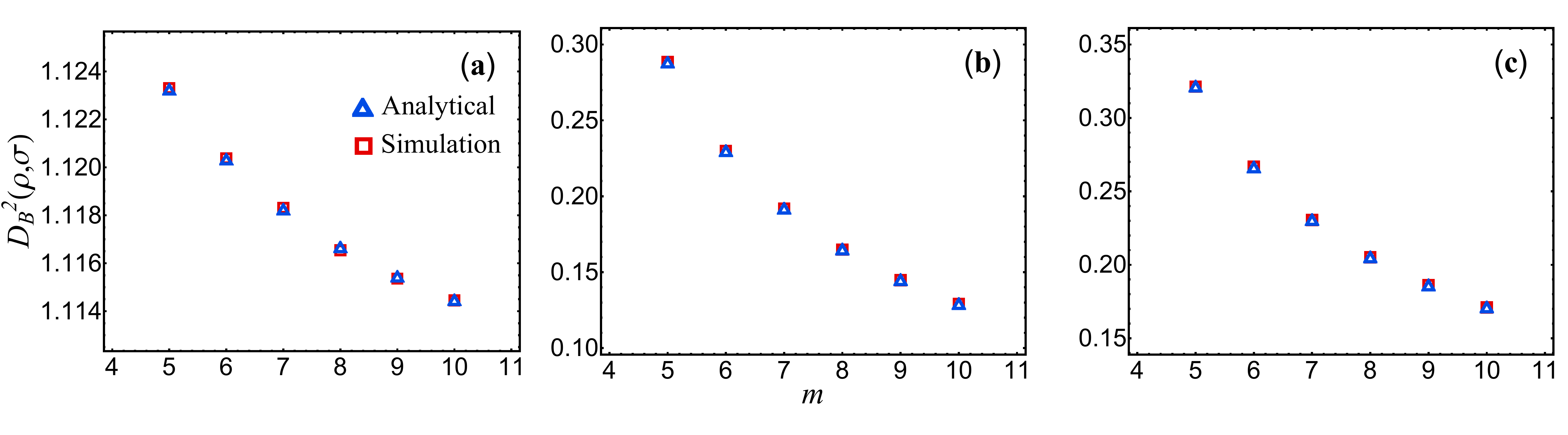}
\centering
\caption{Mean square Bures distance between fixed density matrix $\sigma$ and random density matrix $\rho$ for various choices of $\sigma$: (a) a pure state (b) maximally mixed state (c) having eigenvalues (0.09,0.12,0.21,0.28,0.30). The dimension of two density matrices is set to be $n=5$ in all cases and dimension $m$ of the auxiliary subsystem associated with $\rho$ varies from $n$ to $n+5$.}
\label{fig3}
\end{figure*}
In Fig.~\ref{fig1} we show the plots of random matrix based analytical results for the spectral density of $\tau$ for different combinations of $n, m$ values and three choices of the fixed density matrix $\sigma$. Figure~\ref{fig2} displays the results when the fixed density matrix corresponds to a pure state or a maximally mixed state. For the former, as indicated above, the only nonzero eigenvalue of the product matrix $\tau$ coincides with the fidelity, for which the PDF has been plotted. For the case of $\sigma$ a maximally mixed state, spectral density of $\tau$ has been shown. In all these figures, along with the analytical predictions, Monte-Carlo based simulation results have been also displayed, and we find very good agreements.

\subsubsection{Mean root fidelity and mean square Bures distance}

The mean of the square root of fidelity, $\langle\sqrt{\mathcal{F}(\rho,\sigma)}\,\rangle$, can be calculated as
\begin{equation}
\label{fidla}
\langle\sqrt{\mathcal{F}}\,\rangle=\langle\tr(\sqrt{\tau})\rangle=\expval{ \sum_{j=1}^{n}\lambda_{j}^{1/2} }.
\end{equation}
This can be evaluated using the spectral density $p_{\lambda}(\lambda)$ as
\begin{equation}
\label{fidlb}
\langle\sqrt{\mathcal{F}}\,\rangle=n\int_{-\infty}^{\infty}\lambda^{1/2} \,p_{\lambda}(\lambda) d\lambda.
\end{equation}
Now, either we can use the final expression of $p_{\lambda}(\lambda)$ to evaluate the above, or we can use the Laplace-inverse form as in Eq.~\eqref{pl1}. The latter enables us to express the desired result in terms of the average of $x^{1/2}$ with respect to the density $p_x(x)$,
\begin{align}
\label{fidlc}
\nonumber
&\langle\sqrt{\mathcal{F}}\,\rangle=\mathcal{L}^{-1}\Big[n\Gamma(nm) s^{1-nm}~\int_{-\infty}^{\infty}\!\!\lambda^{1/2} p_{x}(s\lambda)d\lambda\Big]_{t=1}\\
\nonumber
&=\mathcal{L}^{-1}\left[n\Gamma(nm) s^{-nm-1/2}\right]_{t=1}\int_{-\infty}^\infty x^{1/2} p_x(x)dx\\
&=\frac{n}{(nm)_{1/2}}\int_{-\infty}^\infty x^{1/2} p_x(x)dx,
\end{align}
 where $(\alpha)_\beta=\Gamma(\alpha+\beta)/\Gamma(\alpha)$ is the Pochhammer symbol.
As shown in the Appendix~\ref{AppB}, both approaches leads to the final expression of the mean root fidelity as 
\begin{align}
\label{mrf1}
\langle\sqrt{\mathcal{F}}\,\rangle&=\frac{1}{(nm)_{1/2}}\prod_{r>l}\frac{1}{(a_r-a_l)}\sum_{i=1}^{n}\det[\xi^{(i)}_{jk}]_{j,k=1}^n,
\end{align}
where 
\begin{align}
\label{xiijk}
\xi^{(i)}_{jk}=\begin{cases}(m-i+1)_{1/2}\,a_j^{i-3/2}, &  k=i,\\
 a_{j}^{k-1}, &  k\ne i, 
\end{cases}
\end{align}
The mean square Bures distance $D_B^2(\rho,\sigma)$ then readily follows using Eq.~\eqref{DB2}.

Again, when $\sigma$ is either a pure state or a maximally mixed state, the above results can be evaluated in a limiting manner. However, similar to the spectral density of $\tau$, it is easier to handle these extremes directly, as done below.\\

\noindent
\emph{Pure state $\sigma$}: The mean root fidelity in this case be immediately evaluated with the help of Eq.~\eqref{pFpure} as
\begin{equation}
\langle\sqrt{\mathcal{F}}\,\rangle
=\frac{(m)_{1/2}}{(nm)_{1/2}}.
\end{equation}

\noindent
\emph{Maximally mixed state $\sigma$}:
To evaluate mean root fidelity in this case we utilize Eq.~\eqref{fidlc}. Explicit results for eigenvalue-moments of Wishart-Laguerre ensemble is known in the existing literature~\cite{MS2011}, which can be used to obtain
\begin{equation}
\langle\sqrt{\mathcal{F}}\,\rangle=\frac{2}{\sqrt{n}\,(nm)_{1/2}}\sum_{i=1}^n\binom{1/2}{i}\binom{1/2}{i-1}\frac{(m)_{3/2-i}}{(n+1)_{-i}},
\end{equation}
where $\binom{\alpha}{\beta}$ is the binomial coefficient.

We corroborate the ensuing analytical results for mean square Bures distance $D_B^2(\rho,\sigma)$ between a fixed density matrix $\sigma$ and a random density matrix $\rho$ by comparing them with Monte Carlo simulation datasets. These are shown in Fig.~\ref{fig3} for three choices of $\sigma$. We can observe very good agreement between our analytical and simulated results.

\begin{figure*}[!t]
\includegraphics[width=1.0\linewidth]{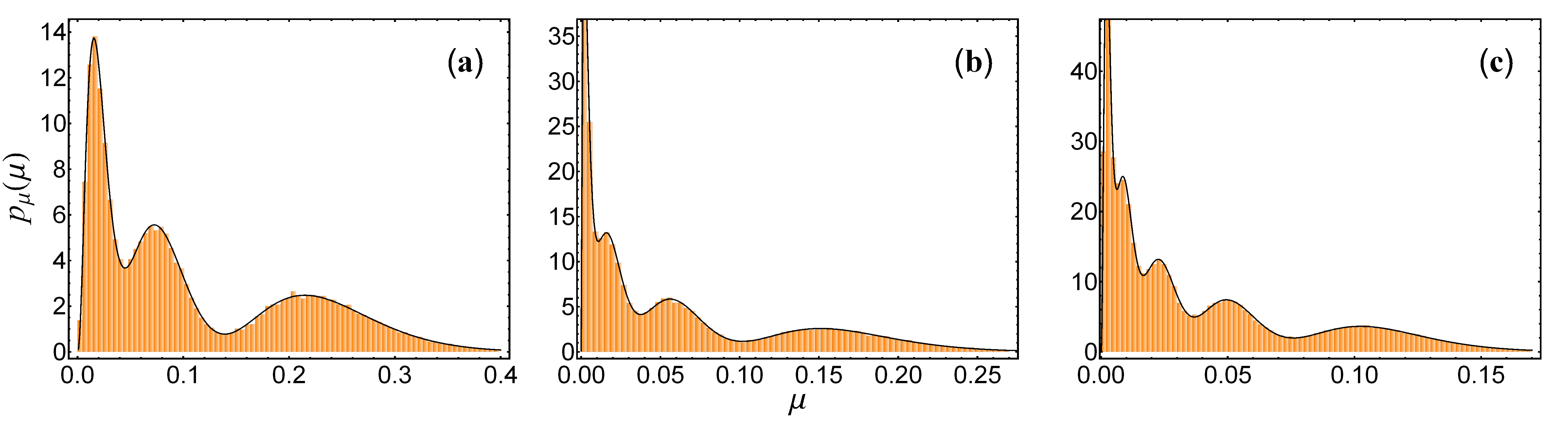}
\centering
\caption{Plots of spectral density of the matrix $\chi=\sqrt{\rho_1}\rho_2 \sqrt{\rho_1}$. The dimensions $(n,m_1,m_2)$ are (a) $(3,6,7)$, (b) $(4,5,8)$, and (c) $(5,8,10)$. The solid curves represents analytical results and the histograms are based on Monte Carlo simulations.}
\label{fig4}
\end{figure*}

\subsection{Two random density matrices}

Let us consider two independent random density matrices $\rho_1$ and $\rho_2$ taken from the distribution given in Eq.~\eqref{Prho}, but with distinct $m$ values in general. We calculate below the mean root fidelity between these two density matrices which then readily leads to the corresponding mean square Bures distance.

\subsubsection{Spectral density of $\chi=\sqrt{\rho_1}\rho_2 \sqrt{\rho_1}$:}

In this case, two random density matrices are involved. Therefore, PDF of the matrix $\chi$ can be obtained as
\begin{equation}
\label{pchia}
\mathcal{P}_\chi(\chi)= \int\int \delta(\chi -\sqrt{\rho_1}\rho_2 \sqrt{\rho_1})~\mathcal{P}_{1}(\rho_1)~\mathcal{P}_{2}(\rho_2)~d[\rho_1]~d[\rho_2],
\end{equation}
where $\mathcal{P}_{1}(\rho_1)$ and $\mathcal{P}_{2}(\rho_2)$ are as in Eq.~\eqref{Prho} but with $m=m_1$ and $m_2$, respectively. As shown in the Appendix~\ref{AppA}, the above can be mapped to a dual inverse-Laplace transform of integrals performed with respect to two independent Wishart-Laguerre random matrices, i.e.,
\begin{align}
\label{pchib}
\nonumber
\mathcal{P}_\chi(\chi)&=\mathcal{L}^{-1}\Big[\Gamma(nm_1)\Gamma(nm_2) s_{1}^{-nm_1}s_{2}^{-nm_2}\\
\nonumber
&\times\int\int\delta(\chi -s_{1}^{-1}s_{2}^{-1}\sqrt{W_1}W_2 \sqrt{W_1})\\
&\times P_1(W_1)~P_2(W_2)~d[W_1]~d[W_2]\Big]_{t_1=t_2=1}.
\end{align}
According to this equation, we can calculate the statistics of $\chi$ by using the statistics of $\sqrt{W_1}W_2 \sqrt{W_1}$. For instance, the marginal density $p_\mu(\mu)$ for a generic eigenvalue $\mu$ of the matrix $\chi$ can be written in terms of the marginal density $p_y(y)$ of a generic eigenvalue $y$ of the matrix $\sqrt{W_1}W_2 \sqrt{W_1}$ as
\begin{align}
\label{pmua}
p_{\mu}(\mu)=\mathcal{L}^{-1}\big[\Gamma(nm_1)\Gamma(nm_2) s_{1}^{1-nm_1}s_{2}^{1-nm_2}
p_{y}(s_1 s_2\mu)\big]_{t_1= t_2=1}.
\end{align}
The spectral density $p_y(y)$ of the matrix $\sqrt{W_1}W_2 \sqrt{W_1}$ follows from the result for product of two Wishart-Laguerre matrices. It is given in terms of Meijer G-functions as~\cite{AIK2013},
\begin{align}
\label{py}
\nonumber
&p_{y}(y)=\frac{1}{n}\sum_{j=0}^{n-1} G_{1,3}^{1,0} 
\left(
\begin{matrix}
j+1\\
0; -v_{1},-v_{2}
\end{matrix} 
\bigg| y\right) 
G_{1,3}^{2,1} 
\left(
\begin{matrix}
-j\\
v_{1},v_{2};0
\end{matrix} 
\bigg| y \right)\Theta(y)~~\\
\nonumber
&=\frac{1}{n}\sum_{j=0}^{n-1}\sum_{k=0}^{j}\frac{(-y)^{k}}{k!(k+v_1)!(k+v_2)!(j-k)!}
G_{1,3}^{2,1} 
\left(
\begin{matrix}
-j\\
v_{1},v_{2};0
\end{matrix} 
\bigg| y \right)\\
&\hspace{6cm}\times\Theta(y).
\end{align}
where $v_1=m_1-n$ and $v_2=m_2-n$. In the second line of the above equation, the first Meijer G-function has been expanded as a polynomial~\cite{AIK2013}. The dual Laplace inversion over $s_1$ and $s_2$ can be performed using results given in Ref.~\cite{PBM1992}; see Appendix~\ref{AppC}. The expression for the marginal density of a generic eigenvalue of the matrix $\chi$ then follows as
\begin{align}
\label{pmub}
\nonumber
p_{\mu}(\mu)&=\frac{\Gamma(nm_1)\Gamma(nm_2)}{n}\sum_{j=0}^{n-1}\sum_{k=0}^{j}\frac{(-\mu)^k}{k!(k+v_1)!(k+v_2)!(j-k)!}\\
& \times G_{3,3}^{2,1} 
\left(
\begin{matrix}
-j;~nm_1-k-1, nm_2-k-1\\
v_{1},v_{2};0
\end{matrix} 
\bigg| \mu \right)\Theta(\mu).
\end{align}
We compare this analytical result with numerical simulation involving an ensemble of two independent random density matrices which are constructed using Eq.~\eqref{map}. These plots, for various combinations of the dimensions $n,m_1,m_2$, are shown in Fig.~\ref{fig4}. Here also, we find an impressive agreement between analytical and numerical results.

\subsubsection{Mean root fidelity}

The mean root fidelity $\langle\sqrt{\mathcal{F}(\rho_1,\rho_2)}\,\rangle$ can be evaluated using the spectral density $p_{\mu}(\mu)$ as
\begin{equation}
\label{fidc}
\langle\sqrt{\mathcal{F}}\,\rangle=n\int_{-\infty}^{\infty}\mu^{1/2} \,p_{\mu}(\mu) d\mu.
\end{equation}
Now, similar to the earlier calculation, either we can use $p_{\mu}(\mu)$ from Eq.~\eqref{pmub} to evaluate the above average, or we can use the Laplace-inverse form as in Eq.~\eqref{pmua}. The latter enables us to express the desired result in terms of the average of $y^{1/2}$ with respect to the density $p_y(y)$,
\begin{align}
\label{fidd}
\langle\sqrt{\mathcal{F}}\,\rangle
=\frac{n}{(nm_1)_{1/2}(nm_2)_{1/2}}\int_{-\infty}^{\infty}y^{1/2} p_y(y)dy.
\end{align}
As shown in Appendix~\ref{AppB} both approaches give the following result after simplification,
\begin{align}
\label{mrf2}
\nonumber
&\langle\sqrt{\mathcal{F}}\,\rangle=\frac{2}{(nm_1)_{1/2}(nm_2)_{1/2}}\\
&\times\sum_{k=1}^{n}\frac{(-1)^{n-k}(k)_{1/2}(k+v_1)_{1/2}(k+v_2)_{1/2}}{\Gamma(n-k+1)\,\Gamma(k-n+1/2)}.
\end{align}
This expression generalizes the explicit formulas provided by Zyczkowski and Sommers for $n=2,3$ and arbitrary $m_1=m_2$~\cite{ZS2005}.
The mean square Bures distance $D_B^2(\rho_1,\rho_2)$ follows from Eq.~\eqref{mrf2} using Eq.~\eqref{DB2}.

\begin{figure}[!t]
\includegraphics[width=0.9\linewidth]{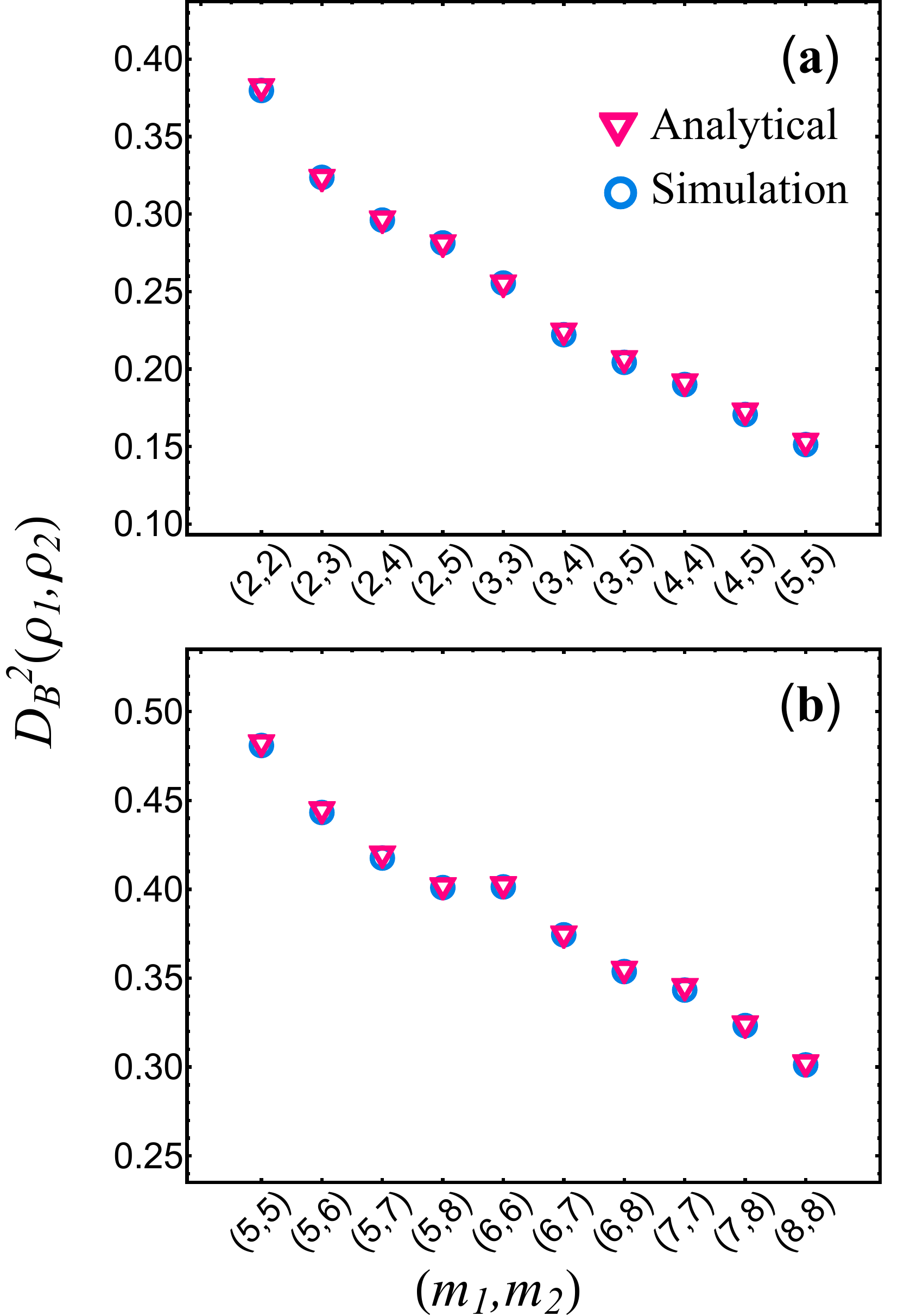}
\centering
\caption{Mean square Bures distance between two independent random density matrices with (a) $n=2$ and (b) $n=5$. In both cases various combination of $m_1, m_2$ values have been considered as depicted on the horizontal axes.}
\label{fig5}
\end{figure}

The analytical prediction for the mean square Bures distance between two random density matrices is contrasted with Monte-Carlo based numerical simulations in Fig.~\ref{fig5}. We again find very good agreements which validate our analytical results.

\section{Comparison with coupled kicked tops simulation}
\label{SecKT}

Coupled quantum kicked tops systems have been used extensively by researchers to study various aspects related to the random bipartite systems such as entanglement, effect of chaos, and even distance between states~\cite{MS1999,BL2002,FMT2003,BL2004,DK2004,TMD2008,KAT2013,K2020,SK2021,PPZ2016}. In this section, we compare our analytical results for the mean square Bures distance with those evaluated using random density matrices generated in coupled kicked tops simulations. 

The Hamiltonian for the coupled kicked tops (CKT) system is given by~\cite{MS1999,BL2002},
\begin{equation}
\label{ham}
H=H_1\otimes\mathds{1}_{m}+\mathds{1}_{n}\otimes H_2+H_{12}.
\end{equation}
Here,
\begin{equation}
\label{indham}
H_r=\frac{\pi}{2}J_{y_r}+\frac{\kappa_r}{2j_r}J_{z_r}^{2}\sum_{\nu=-\infty}^{\infty}\delta(t-\nu), r=1,2
\end{equation}
represent the Hamiltonians for the individual tops~\cite{HKS1987,H2010} and the interaction term $H_{12}$ is of the form
\begin{equation}
\label{intr}
H_{12}=\frac{\epsilon}{\sqrt{j_{1}j_{2}}}(J_{z1}\otimes J_{z2})\sum_{\nu=-\infty}^{\infty}\delta(t-\nu).
\end{equation}
The Hamiltonians $H_1$ and $H_2$ are associated with $n (=2j_1+1)$-dimensional and $m (=2j_2+1)$-dimensional Hilbert spaces $\mathcal{H}^{(n)}$ and $\mathcal{H}^{(m)}$ respectively. The full Hamiltonian for the coupled kicked tops ($H$) correspond to the Hilbert space $\mathcal{H}^{(nm)}=\mathcal{H}^{(n)}\otimes \mathcal{H}^{(m)}$. The first term in $H_r$ gives the free precession of the $r$th top around $y$ axis with an angular frequency $\pi/2$ and the second term signifies periodic $\delta$-function kicks. The angular momentum operators for the $r$th top are $J_{x_r},J_{y_r},J_{z_r}$, and $j_r$ is the quantum number associated with the operator $J_{r}^{2}=J_{x_r}^2+J_{y_r}^2+J_{z_r}^2$. Further, $\kappa_1$ and $\kappa_2$ are the stochasticity parameters which relate to the kick strengths and control the chaotic behaviour of the tops. The coupling between the two tops is provided by the parameter $\epsilon$.

\begin{figure}[!t]
\includegraphics[width=0.9\linewidth]{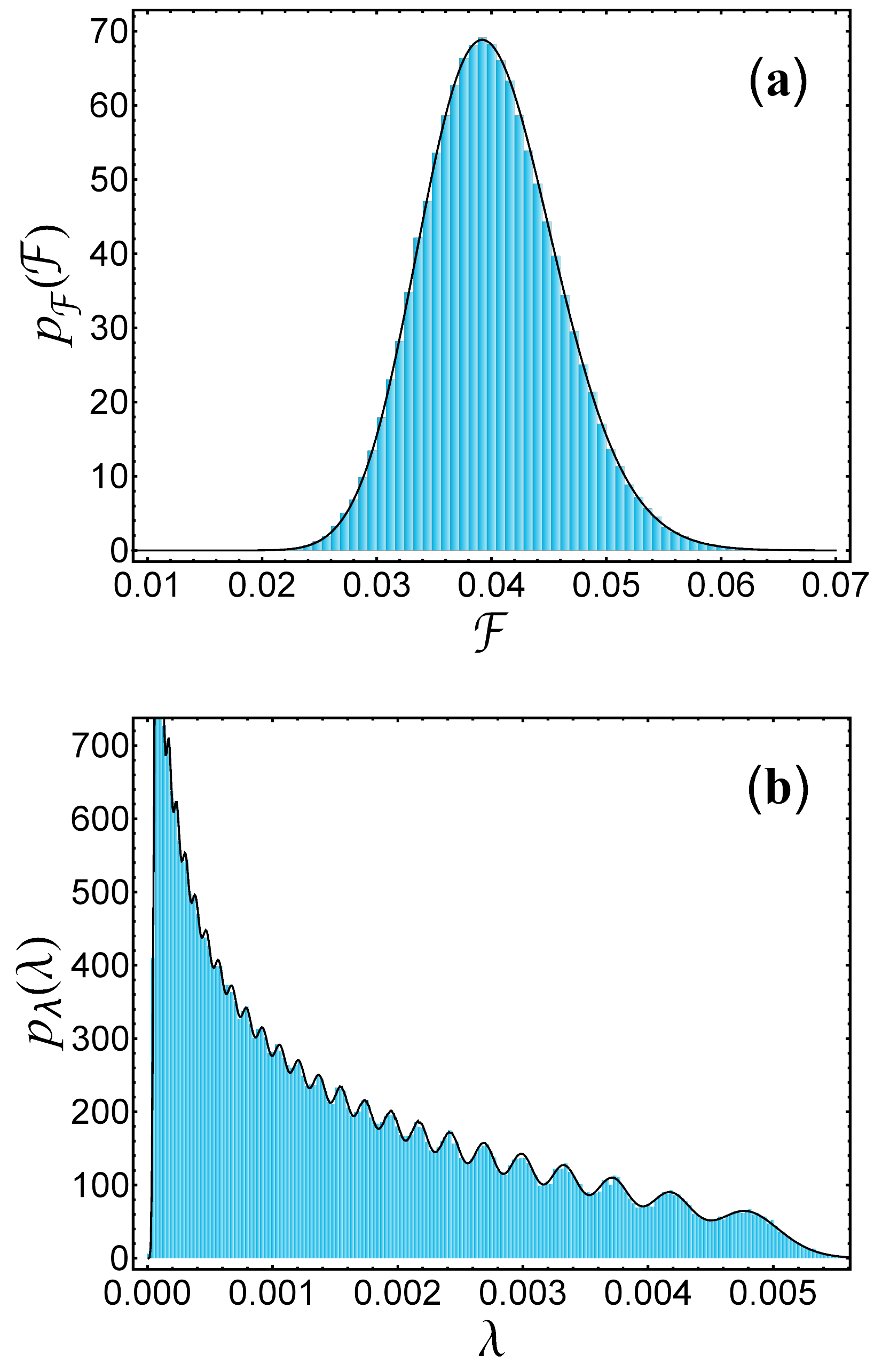}
\centering
\caption{Comparison between random matrix theory (RMT) based analytical prediction (solid curve) and coupled kicked tops (CKT) simulation result (histogram) for (a) the PDF of fidelity between a random density matrix $\rho$ and a pure state $\sigma$, (b) spectral density of the matrix $\tau=\sqrt{\sigma}\rho\sqrt{\sigma}$ for $\sigma$ maximally mixed. The dimensions $(n,m)$ associated with the random matrix $\rho$ for these two plots are $(25,45)$ and $(25,35)$, respectively.}
\label{fig6}
\end{figure}
\begin{figure}[!t]
\includegraphics[width=0.9\linewidth]{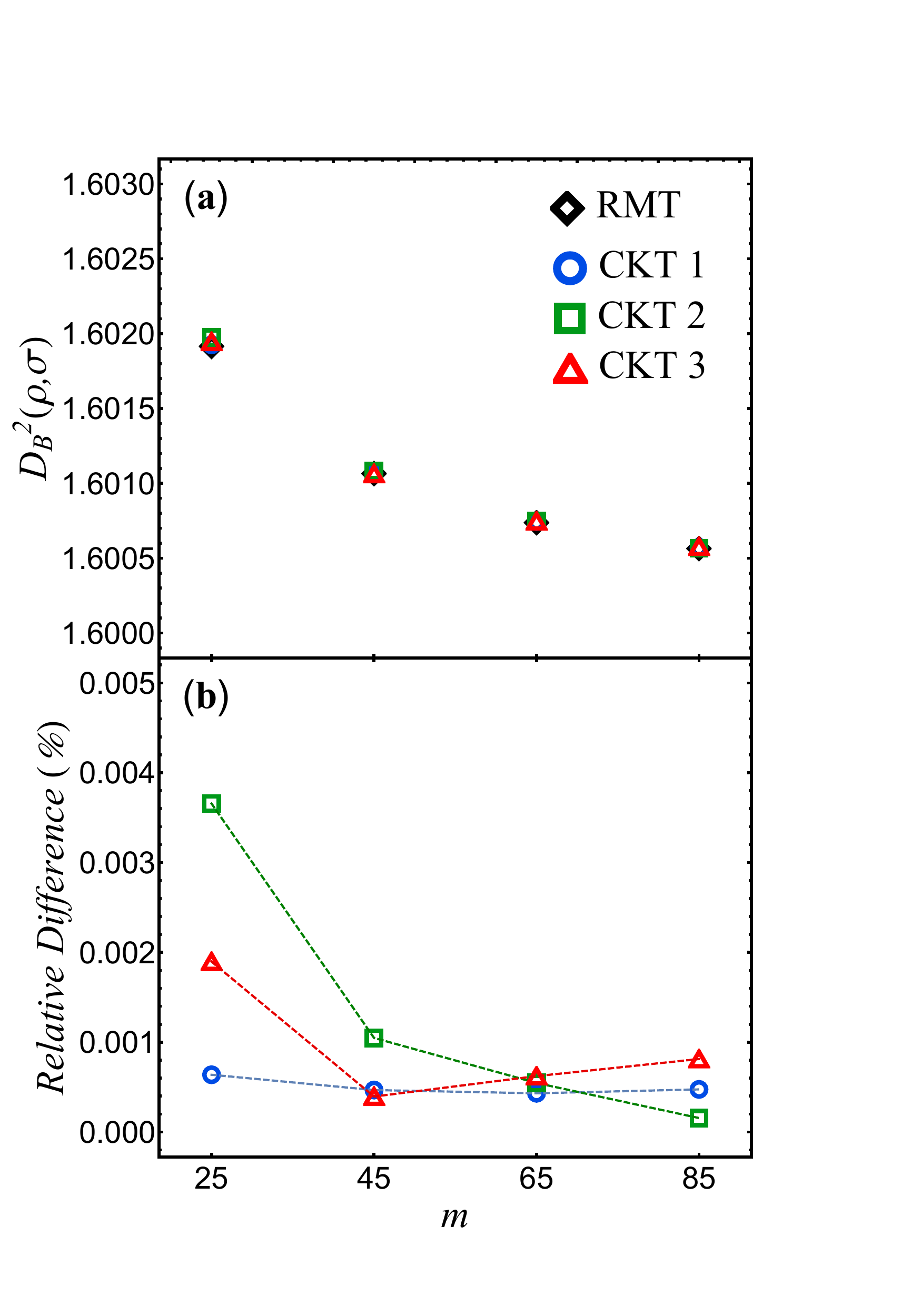}
\centering
\caption{Comparison between RMT analytical and CKT simulation results: (a) mean square Bures distance between random density matrix $\rho$ and pure state $\sigma$, and (b) the corresponding percent relative differences, $100|([D^{2}_{\rho,\sigma}]_{CKT}/[D^{2}_{\rho,\sigma}]_{RMT}-1)|\%$. The dimensions of both density matrices is $n=25$ and the $m$ value for $\rho$ is varied over four values as indicated on the horizontal axis. The CKT simulations have been carried out for three sets of stochasticity and coupling parameters ($\kappa_1, \kappa_2, \epsilon$), viz., CKT 1: ($7,8,1$), CKT 2: ($6,7,0.75$), CKT 3: ($7,9,0.5$).}
\label{fig7}
\end{figure}

\begin{figure}[!t]
\includegraphics[width=0.9\linewidth]{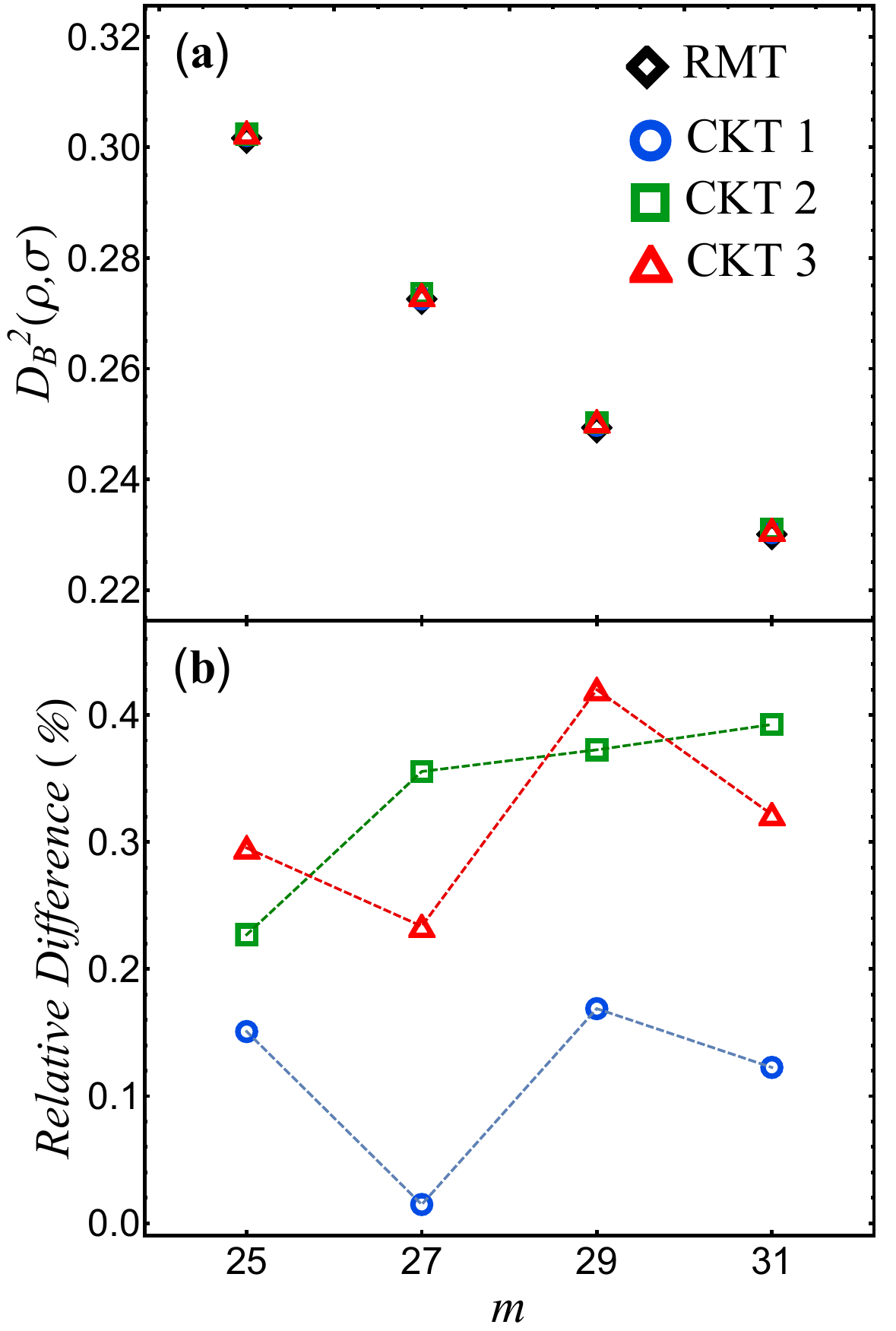}
\centering
\caption{Comparison between RMT analytical and CKT simulation results: (a) mean square Bures distance between random density matrix $\rho$ and maximally mixed state $\sigma$, (b) the corresponding percent relative differences, $100|([D^{2}_{\rho,\sigma}]_{CKT}/[D^{2}_{\rho,\sigma}]_{RMT}-1)|\%$. The dimensions of both density matrices is $n=25$ and the $m$ value for $\rho$ is varied over four values. The CKT simulations have been carried out for three sets of stochasticity and coupling parameters as mentioned in Fig.~\ref{fig7} caption.}
\label{fig8}
\end{figure}

\begin{figure}[!t]
\includegraphics[width=0.9\linewidth]{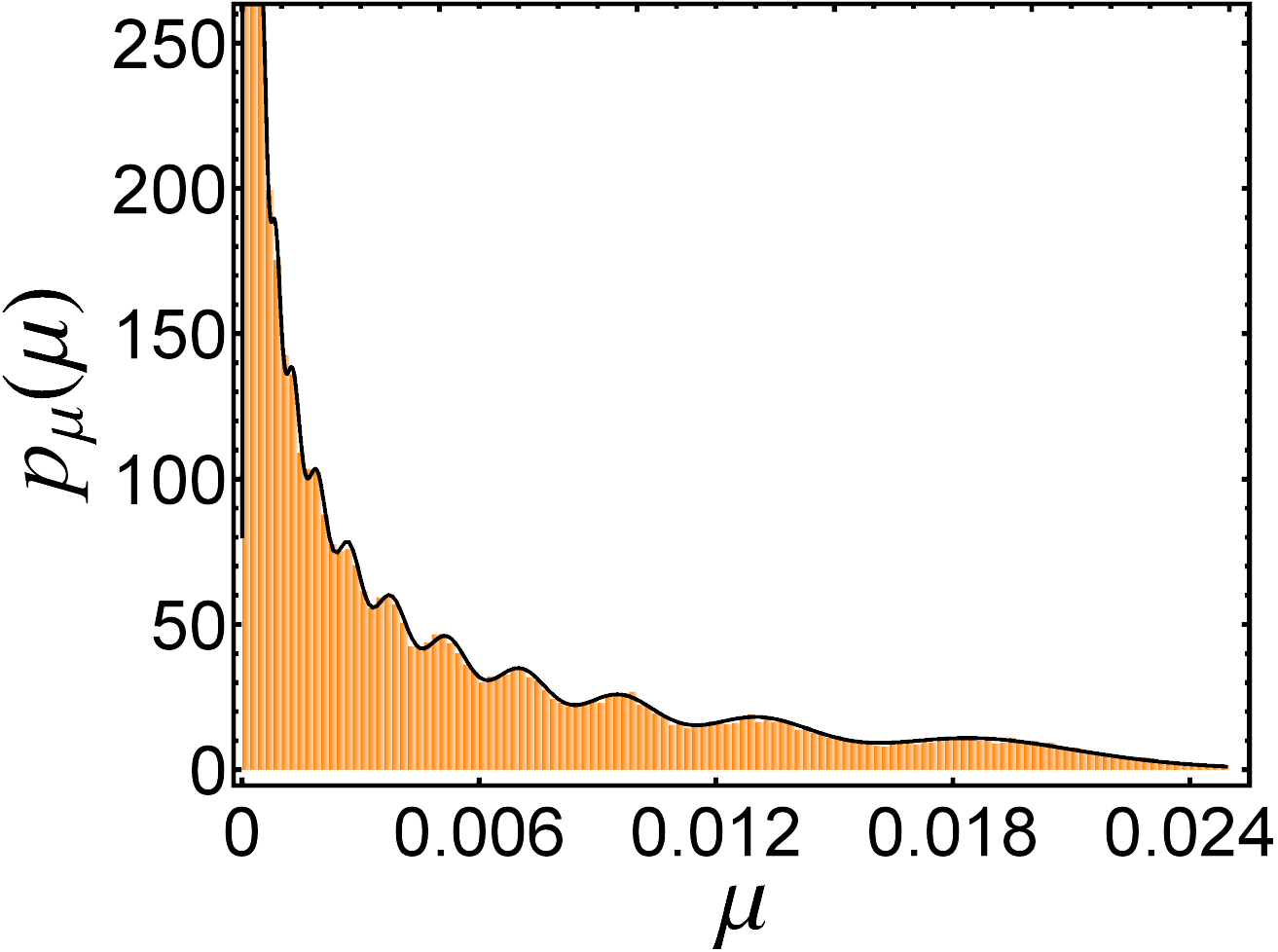}
\centering
\caption{Comparison between RMT based analytical prediction (solid curve) and CKT simulation result (histogram) for the spectral density of matrix $\chi=\sqrt{\rho_1}\rho_2\sqrt{\rho_1}$. The dimensions used are $(n, m_1,m_2)=(15,17,21)$.}
\label{fig9}
\end{figure}

\begin{figure}[!t]
\includegraphics[width=0.9\linewidth]{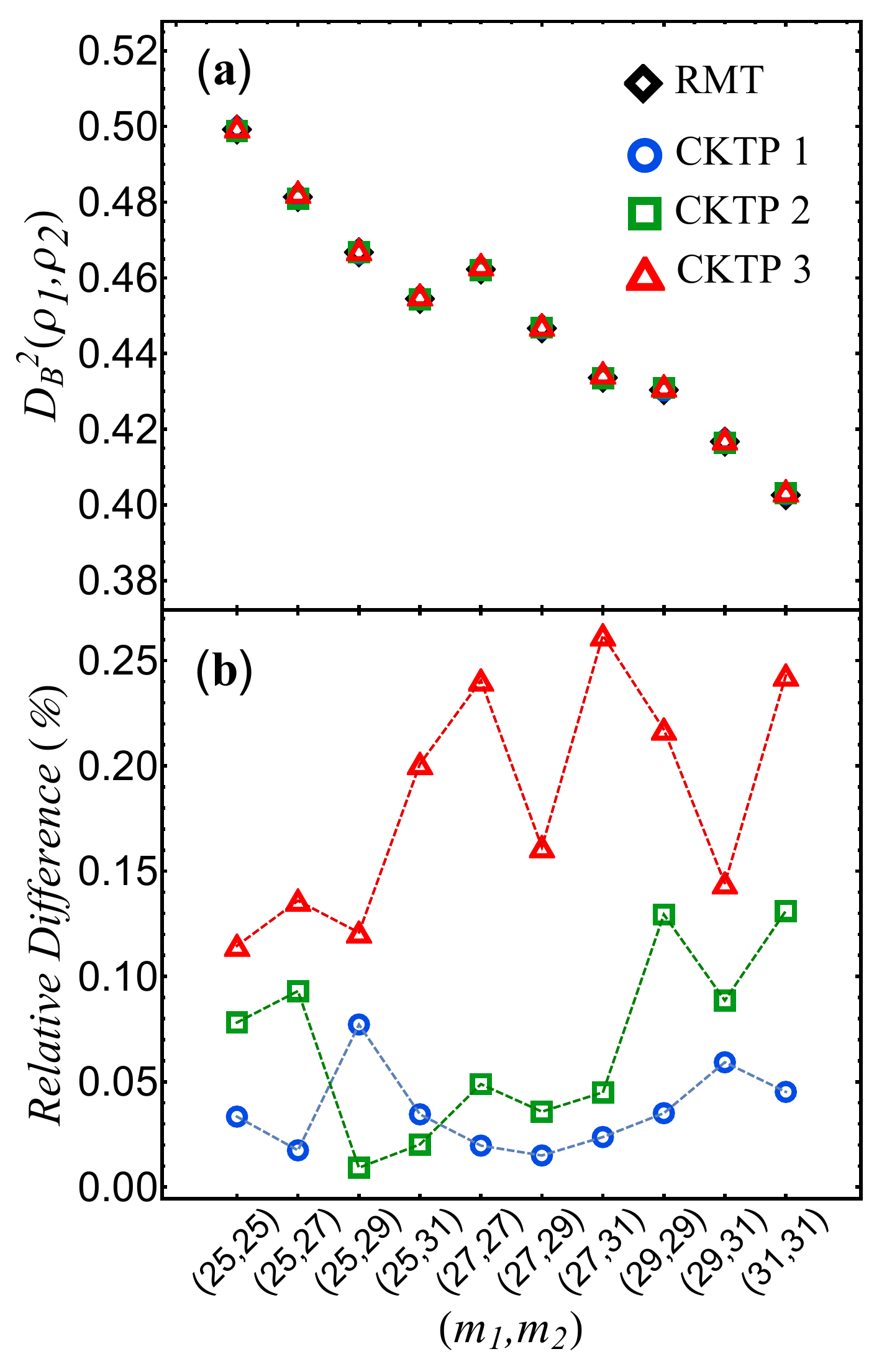}
\centering
\caption{Comparison between RMT analytical and coupled kicked tops pairs (CKTP) simulation results: (a) mean-square Bures distance between two random density matrices $\rho_1, \rho_2$ of dimension $n=25$ and several $m_1$, $m_2$ combinations, and (b) the corresponding percent relative differences. The CKTP simulations have been carried out for three sets of parameters ($\kappa_{1}^{A}, \kappa_{2}^{A}, \epsilon^{A};\kappa_{1}^{B}, \kappa_{2}^{B}, \epsilon^{B}$), viz., CKTP 1: ($8,7,0.5;7,8,1$), CKTP 2: ($6,7,0.8;6,8,0.75$), CKTP 3: ($7,8,0.75;8,7,0.75$).}
\label{fig10}
\end{figure}

We note that the operators for the two independent tops commute and that between the $\delta$-function kicks only free-precession parts of the Hamiltonian survive, whereas at the instants of the kicks they are ineffective. Consequently, the Floquet operator which evolves a state vector from immediately after one kick to immediately after the next can be obtained using the Hamiltonian in Eq.~\eqref{ham} as~\cite{MS1999,BL2002},
\begin{equation}
\label{floq}
U=U_{12}(U_1\otimes U_2),
\end{equation}
where,
\begin{equation}
U_{r}=\mathrm{exp}\left(-\frac{\imath\kappa_r}{2j_r}J_{z_r}^{2}\right)\mathrm{exp}\left(-\frac{\imath\pi}{2}J_{y_r}\right), ~r=1,2,
\end{equation}
\begin{equation}
U_{12}=\mathrm{exp}\left(-\frac{\imath\epsilon}{\sqrt{j_1j_2}}J_{z_1}\otimes J_{z_2}\right),
\end{equation}
with $\imath=\sqrt{-1}$ being the imaginary-number unit. We use the Floquet operator $U$ to implement the iteration scheme $|\psi(\nu)\rangle=U|\psi(\nu-1)\rangle$ and hence obtain a time-evolved state from an initial state $|\psi(0)\rangle$. As in other works~\cite{MS1999,BL2002}, we choose the initial state as the product of directed angular momentum states associated with the two tops. After ignoring a certain number of iterations that fall in the transient regime, we consider the reduced density matrices obtained by partial tracing over one of the tops, viz., $\rho(\nu)=\mathrm{tr}_{m}(|\psi(\nu)\rangle\langle\psi(\nu)|)$. These reduced density matrices are distributed according to the Hilbert-Schmidt probability measure when the stochasticity parameters and coupling parameter are sufficiently large~\cite{BL2002,KSA2017,K2020}.

We simulate $60000$ reduced density matrices using the above procedure and compute the spectral density of the product matrix $\tau=\sqrt{\sigma}\rho\sqrt{\sigma}$, as well as the corresponding mean square Bures distance. For the fixed matrix $\sigma$, we choose a pure state and then a maximally mixed state. These PDF's are shown in Figs.~\ref{fig6} and the mean square Bures distances are illustrated in Figs.~\ref{fig7} and \ref{fig8}. 

To simulate the spectral density of the product matrix $\chi=\sqrt{\rho_1}\rho_2\sqrt{\rho_2}$ and the corresponding mean square Bures distance, we consider two independent coupled kicked tops systems, say $A$ and $B$. This helps us to realize different $m_1=2j_{2}^{A}+1$ and $m_2=2j_{2}^{B}+1$ values. Here $j_{2}^{A}$ and $j_{2}^{B}$ represent the $j_2$ values for the two coupled kicked tops, respectively. The $n$ value is determined by the common Hilbert-space dimension $2j_{1}^{A}+1=2j_{1}^{B}+1$. We show the comparison between the random matrix analytical and kicked tops simulation results for the spectral density of $\chi$ in Fig.~\ref{fig9} and for the mean square Bures distance in Fig.~\ref{fig10}. 

In all these plots we find impressive agreements which are supported by the observation that the percentage relative difference in the case of mean square Bures distance remains below $0.5\%$.

\section{Summary and Outlook}
\label{SecSum}

We considered random density matrices distributed according to the Hilbert-Schmidt probability measure and derived exact analytical results for the mean root fidelity and mean square Bures distances. We examined these average quantities between a fixed density matrix and a random density matrix, and also between two random density matrices. In the course of derivation, we also obtained spectral densities for products of these density matrices. Moreover, we compared our analytical results with the observations in coupled kicked tops systems and found very good agreements. With this work, we accomplish one of the desired tasks outlined in Ref.~\cite{K2020}.

A natural extension of this work would be to go beyond the averages and explore the feasibility of the calculation of exact higher moments and distributions of various distance measures. Besides the Hilbert-Schmidt measure, another important probability measure over the set of mixed states is the Bures-Hall measure~\cite{H1998,ZS2001,SZ2003,SZ2004,BS2001}. In recent times, there has been a renewed interest in its statistical investigation from the random matrix theory point of view~\cite{FK2016,SK2019,SK2021,Wei2020a,Wei2020b}. In this regard, it would be also of interest to obtain exact results for statistics of various distance measures between these Bures-Hall distributed random density matrices.

\acknowledgements
A.L. would like to acknowledge financial support from Shiv Nadar University.

\appendix

\section{Laplace transform for mapping to Wishart-Laguerre ensemble average}
\label{AppA}

We introduce an auxiliary variable $t$ to replace $1$ present inside the delta function in Eq.~\eqref{Prho}, which is then used in Eq.~\eqref{ptaua}. This gives,
\begin{align}
\label{pdf12}
\nonumber
\mathcal{P}_\tau(\tau;t)&= C\int \delta(\tau -\sqrt{\sigma}\rho\sqrt{\sigma})\\
&~~\times(\det \rho)^{m-n}~\delta(\tr \rho-t)~\Theta(\rho)~d[\rho].
\end{align}
We now take the Laplace transform ($t\to s$), which leads us to
\begin{equation}
\label{a1}
\widetilde{\mathcal{P}_\tau}(\tau;s)= C\int \delta(\tau -\sqrt{\sigma}\rho\sqrt{\sigma})(\det \rho)^{m-n}~e^{-s\tr \rho}~\Theta(\rho)~d[\rho].
\end{equation}
Next, we consider the substitution $\rho=W/s$ with $s>0$, which gives $d[\rho]=s^{-n^2}d[W]$ and $\Theta(\rho)=\Theta(W)$.
Therefore, Eq. \eqref{a1} becomes
\begin{align}
\label{a2}
\nonumber
&\widetilde{\mathcal{P}_\tau}(\tau;s)= Cs^{-nm}\int \delta(\tau -s^{-1}\sqrt{\sigma}W\sqrt{\sigma})\\
\nonumber
&~~~~~\times(\det W)^{m-n}e^{-\tr W}\Theta(W)d[W]\\
&=\Gamma(nm)s^{-nm}\int \delta(\tau -s^{-1}\sqrt{\sigma}W\sqrt{\sigma})P(W)d[W],
\end{align}
where we have used Eq.~\eqref{PW} to write the second line. The PDF $\mathcal{P}_{\tau}(\tau)$ can now be obtained by taking the inverse Laplace transform and setting $t=1$, as it appears in Eq.~\eqref{ptaub}.

Similarly, for the case of two random density matrices as in Eq.~\eqref{pchia}, we introduce two auxiliary variables $t_1$ and $t_2$ to replace the 1's within the delta functions in the two densities $\mathcal{P}_1(\rho_1)$ and $\mathcal{P}_2(\rho_2)$, and then consider a dual Laplace transform $t_1\to s_1$ and $t_2\to s_2$. The integrals over the two random density matrices can then be mapped to integrals over two Wishart-Laguerre matrices by the substitutions $\rho_1=W_1/s_1$ and $\rho_2=W_2/s_2$ with $s_1,s_2>0$. The desired expression, Eq.~\eqref{pchib}, then follows by considering dual Laplace inversion with $t_1,t_2$ set equal to 1.

\section{Derivation of mean root fidelity $\langle\sqrt{\mathcal{F}}\rangle$}
\label{AppB}

We consider first the fidelity between one fixed density matrix and one random density matrix. To evaluate mean root fidelity $\langle \sqrt{\mathcal{F}(\rho,\sigma)}\rangle$ using Eq.~\eqref{fidlb}, we use the expression for $p_\lambda(\lambda)$ from Eq.~\eqref{pl2}. Then, we push the $\lambda^{1/2}$ factor inside the determinant to the $k=i$th column of the matrix $[\eta_{jk}^{(i)}]$. Since only this column involves the integration variable $\lambda$, we can perform the integration readily using the Beta-function Euler integral, 
\begin{align}
\int_{-\infty}^\infty \lambda^{\alpha-1} (1-a \lambda)^{\beta-1}\Theta(\lambda)\Theta(1-a\lambda)d\lambda=\frac{\Gamma(\alpha)\Gamma(\beta)}{a^{\alpha}\Gamma(\alpha+\beta)},
\end{align}
with $\alpha,\beta,a>0$. Combining this result with the other pre-existing factors gives the elements of this column as $a_j^{i-3/2}(m-i+1)_{1/2}/\Gamma(n m+1/2)$. The factor $1/\Gamma(n m+1/2)$ is now pulled outside the determinant and the summation, and combined with the already existing factor $\Gamma(nm)$ to give $1/(nm)_{1/2}$. Therefore, finally, the element $\xi_{jk}^{(i)}$ of the matrix within the determinant is given by Eq.~\eqref{xiijk} and the mean root fidelity is obtained as Eq.~\eqref{mrf1}. 

To derive the same result using Eq.~\eqref{fidlc}, we plug the expression for $p_x(x)$ in it and push the factor $x^{1/2}$ in the $k=i$th column of the matrix $[\zeta_{jk}]$ and perform the integration using the Gamma-function Euler integral,
\begin{align}
\int_{-\infty}^\infty x^{\alpha-1}e^{-a x}\Theta(x)dx=\frac{\Gamma(\alpha)}{a^\alpha},
\end{align}
for $\alpha, a>0$. This, when combined with the pre-exising factor, gives at once the matrix element $\xi_{jk}^{(i)}$ as $a_j^{i-3/2}(m-i+1)_{1/2}$. Consequently, we again obtain Eq.~\eqref{mrf1}. 

We now consider the calculation of fidelity between two random density matrices. Let us consider first the expression for the mean fidelity given in Eq.~\eqref{fidd}. The $y$-integral in this equation can be performed using an integral identity of the Meijer G-function~\cite{PBM1990}, which gives
\begin{align}
\nonumber
&\int_{0}^{\infty} y^{k+1/2} G_{1,3}^{2,1} 
\left(
\begin{matrix}
-j\\
v_{1},v_{2};0
\end{matrix} 
\bigg| y \right)dy\\
&=\frac{\Gamma(k+v_1+3/2)\Gamma(k+v_2+3/2)\Gamma(j-k-1/2)}{\Gamma(-k-1/2)}.
\end{align}
It should be noted that the Heaviside theta function $\Theta(y)$ in Eq.~\eqref{py} has restricted the integration domain to $(0,\infty)$. The use of this integral in Eq.~\eqref{fidd} leads to
\begin{align}
\label{ay}
\nonumber
&n\int_{-\infty}^{\infty} y^{1/2} p_y(y)dy 
=\sum_{j=0}^{n-1}\sum_{k=0}^j \frac{(-1)^k}{(j-k)!}\\
&\times\frac{\Gamma(k+v_1+3/2)\Gamma(k+v_2+3/2)\Gamma(j-k-1/2)}{k!(k+v_1)!(k+v_2)!\,\Gamma(-k-1/2)}.
\end{align}
As described in Ref.~\cite{AIK2013}, after reordering of the sums and application of Euler's reflection formula for the Gamma function, the above can be simplified to
\begin{align}
\label{ax}
\nonumber
&n\int_{-\infty}^{\infty} y^{1/2} p_y(y)dy \\
\nonumber
&=\sum_{k=0}^{n-1}\frac{\Gamma(k+3/2)\Gamma(k+v_1+3/2)\Gamma(k+v_2+3/2)}{k!(k+v_1)!(k+v_2)!}\\
&\times \sum_{j=0}^{n-k-1}\frac{(-1)^j}{j!\, \Gamma(3/2-j)}.
\end{align}
Now, we note that the summand of the sum in the third line can be written as $(-1)^j/[j!\, \Gamma(3/2-j)]=[1/\Gamma(3/2)](-1)^j\binom{1/2}{j}$. Therefore, using the Binomial series relation~\cite{AIK2013},
\begin{equation}
\sum_{j=0}^{n-k-1}(-1)^j\binom{1/2}{j}=(-1)^{n-k-1}\binom{-1/2}{n-k-1},
\end{equation}
it can be evaluated to a closed form. We use this in Eq.~\eqref{ax}, along with a shift $k\to k-1$ in the summand, and express ratio of two Gamma functions using Pochhammer symbol. This gives
\begin{align}
\nonumber
&n\int_{-\infty}^{\infty} y^{1/2} p_y(y)dy \\
&=\sum_{k=1}^{n}\frac{2\,(-1)^{n-k}(k)_{1/2}(k+v_1)_{1/2}(k+v_2)_{1/2}}{\Gamma(n-k+1)\Gamma(k-n+1/2)},
\end{align}
and consequently we obtain Eq.~\eqref{mrf2}.

If we instead start from Eq.~\eqref{fidc}, using expression for $p_\mu(\mu)$ gives for the $\mu$-integral~\cite{PBM1990},
\begin{align}
\nonumber
&\int_{0}^{\infty} \mu^{k+1/2} G_{3,3}^{2,1} 
\left(
\begin{matrix}
-j;~nm_1-k-1, nm_2-k-1\\
v_{1},v_{2};0
\end{matrix} 
\bigg| \mu \right)d\mu\\
&=\frac{\Gamma(k+v_1+3/2)\Gamma(k+v_2+3/2)\Gamma(j-k-1/2)}{\Gamma(-k-1/2)\Gamma(nm_1+1/2)\Gamma(nm_2+1/2)},
\end{align}
so that
\begin{align}
\nonumber
&n\int_{-\infty}^{\infty} \mu^{1/2} p_\mu(\mu)d\mu\\
\nonumber
&=\frac{1}{(nm_1)_{1/2}(nm_2)_{1/2}}\sum_{j=0}^{n-1}\sum_{k=0}^j \frac{(-1)^k}{(j-k)!}\\
\nonumber
&\times\frac{\Gamma(k+v_1+3/2)\Gamma(k+v_2+3/2)\Gamma(j-k-1/2)}{k!(k+v_1)!(k+v_2)!\,\Gamma(-k-1/2)}\\
&=\frac{n}{(nm_1)_{1/2}(nm_2)_{1/2}}\int_{0}^{\infty} y^{1/2} p_y(y)dy.
\end{align}
Here, the last step follows using Eq.~\eqref{ay}. This matches with Eq.~\eqref{fidd} and therefore leads again to Eq.~\eqref{mrf2}.

\section{Inverse Laplace transform of Meijer G}
\label{AppC}

With the help of the general expression given in Ref.~\cite{PBM1992}, we find the following result for Laplace inversion ($s\to t$),
\begin{align}
\nonumber
\mathcal{L}^{-1}\left[ s^{-\gamma}G^{j,k}_{l,r} 
\left(
\begin{matrix}
\alpha_1,...,\alpha_k;\alpha_{k+1},...,\alpha_l\\
\beta_1,...,\beta_j;\beta_{j+1},...,\beta_r
\end{matrix} 
\bigg| s\mu \right)\right]_t\\
=t^{\gamma-1}G^{j,k}_{l+1,r} 
\left(
\begin{matrix}
\alpha_1,...,\alpha_k;\alpha_{k+1},...,\alpha_l,\gamma\\
\beta_1,...,\beta_j;\beta_{j+1},...,\beta_r
\end{matrix} 
\bigg| \frac{\mu}{t} \right).
\end{align}
We substitute Eq.~\eqref{py} in Eq.~\eqref{pmua} and focus on the portion of the full expression where we have to apply the dual Laplace inversion ($s_1\to t_1, s_2\to t_2$). We use the above result twice to perform the two Laplace inversions successively. We obtain,
\begin{align}
\nonumber
&\mathcal{L}^{-1}\left[ s_1^{-(nm_1-k-1)}s_2^{-(nm_2-k-1)}G^{2,1}_{1,3} 
\left(
\begin{matrix}
-j\\
v_1,v_2;0
\end{matrix} 
\bigg| s_1 s_2\mu \right)\right]_{t_1,t_2}\\
\nonumber
&=t_1^{(nm_1-k-2)}t_2^{(nm_2-k-2)}\\
&\times G^{2,1}_{3,3} 
\left(
\begin{matrix}
-j;nm_1-k-1,nm_2-k-1\\
v_1,v_2;0
\end{matrix} 
\bigg| \frac{\mu}{t_1 t_2} \right).
\end{align}
This result with $t_1,t_2$ set equal to 1 yields Eq.~\eqref{pmub}.


\end{document}